\documentclass[paper=a4,fontsize=12pt]{scrartcl}

\usepackage[version=3]{mhchem} 

\usepackage{graphicx}

\usepackage[ansinew]{inputenc}

\usepackage{textcomp}

\usepackage{upgreek}

\usepackage[T1]{fontenc}

\usepackage[english]{babel}
\usepackage[babel, english=british]{csquotes} 

\usepackage[y2k,alphabetic,nobysame,initials,short-months,short-publishers]{amsrefsH}
\usepackage{amsmath}

\begin{document}

\begin{center}
	\textbf{Enhanced Adhesion of \textit{S.\,Mutans} to Hydroxyapatite after~Inoculation~in~Saliva}
\end{center}

\vspace{2cm}

\noindent  Christian Spengler(1), Nicolas Thewes(1), Friederike Nolle(1), Thomas Faidt(1), Natalia Umanskaya(2), Matthias Hannig(2), Markus Bischoff(3), Karin Jacobs(1)*

\vspace{2cm}

\noindent 1) Experimental Physics, Saarland University, Saarbrücken, Germany

\noindent 2) Clinic of Operative Dentistry, Periodontology and Preventive Dentistry, Saarland University, Homburg, Germany

\noindent 3) Institute of Medical Microbiology and Hygiene, Saarland University, Homburg, Germany

\vspace{5cm}

\noindent *Corresponding author: Karin Jacobs, 

\noindent email: k.jacobs@physik.uni-saarland.de, 

\noindent phone:+49 (0) 681 302 788, 

\noindent fax: +49 (0) 681 302 700

\newpage
\begin{abstract}
\textit{Streptococcus mutans} cells form robust biofilms on human teeth and are strongly related to caries incidents. Hence, understanding the adhesion of \textit{S.\,mutans} in the human oral cavity is of major interest for preventive dentistry. In this study, we report on AFM-based single cell force spectroscopy measurements of \textit{S.\,mutans} cells to hydroxyapatite surfaces. We observe a significant increase of adhesion strength if \textit{S.\,mutans} cells were exposed to human saliva before adhesion. In contrast, \textit{Staphylococcus\,carnosus} cells are almost unaffected by the pre-treatment. Our results demonstrate that \textit{S.\,mutans} cells are well-adapted to their natural environment, the oral cavity. This ability promotes the biofilm-forming capability of that species and hence the production of caries-provoking acids. In consequence, understanding the fundamentals of this mechanism may pave a way towards more effective caries-reducing techniques.  \\

\vspace{2cm}

\noindent Keywords\\

\noindent single-cell force spectroscopy, hydroxyapatite, \textit{Streptococcus mutans}, saliva

\end{abstract}
\newpage

\section*{Introduction}

It is known for decades that \textit{Streptococcus\,mutans} is very closely related to the development of caries and other diseases in the oral cavity\cite{Hamada:1980gc,Loesche:1986vxba,Mitchell:2003ibba}. Furthermore,  it can also enter the bloodstream through wounds in the oral cavity and travel from there through the body and even reach the coronary artery, where it can cause severe cardio-vascular diseases\cite{Abranches:2009fgba}. The main pathogenicity of this organism arises when the cell adheres to appropriate surfaces, since with this step, the formation of a biofilm is initiated. 

The process of caries formation is thereby influenced by substratum (e.\,g. enamel, fluoridated or not) and saliva\cite{Selwitz:2008wtba,Takahashi:2008gpba,Filoche:2010ctba,Loskill:2013jeba}. 
On exposure to saliva, a proteinaceous surface coating -- called pellicle -- is formed almost immediately on all solid substrates\cite{Hannig:2009ba}.This conditioning layer changes the properties of the substrate.
The nature of the chemical groups exposed at the surface mainly define the adhesion forces\cite{Loskill:2013jeba} .

Most studies focus on the adhesion of oral bacteria to salivary agglutinin (SAG) which is adsorbed to the oral pellicle on tooth surfaces \cite{Ray:1999vqbaca,Yoshida:2005caba,Miller:2015crba}. Additionally, it has been shown that the characteristics of biofilm formation by \textit{S.\,mutans} depend on many parameters like for example oxygen content or the presence of specific enzymes in the bacterial cell\cite{Ahn:2007uzbaca,Ahn:2008vjbaca}. Next to other constituents, the salivary sucrose content increases adhesion to surfaces significantly and is also a key factor in the production of biofilms\cite{Costa:2013tabaca}. Furthermore, by using genetically modified \textit{S.\,mutans} cells, the function of many proteins in adhesion processes and biofilm formation on SAG was identified\cite{Ray:1999vqbaca,Yoshida:2005caba,Miller:2015crba}.

Atomic force microscopy(AFM)-based force spectroscopy offers a unique tool to quantitatively investigate crucial parameters of initial bacterial adhesion. By using this technique, cantilevers functionalized with specific proteins of the bacterial cell wall  provide access to probe molecular interactions between these proteins and various substrate surfaces. For example, the binding between SAG and the P1\,adhesin of \textit{S.\,mutans}, which is crucial for adhesion, has been characterized and quantified\cite{Brady:1992wxba,Sullan:2015dtbacada}. 

For this AFM-based force spectroscopy study, we prepared cantilevers with single, viable bacterial cells to probe the interaction of the entire bacterial cell with the substratum dependent on a  pretreatment of the cell\cite{Thewes:2015heba}. Thereby, substrate parameters are kept constant. For force spectroscopy, a controlled, low roughness of the substratum is a precondition, since on rough, natural teeth surfaces, the contact area between bacterial cell and surface is unpredictable. Therefore, as a model tooth material with low roughness and still high biological relevance, we used pressed, sintered and polished high-density pellets of hydroxyapatite (HAP), which is the mineral component of human tooth enamel\cite{Loskill:2013jeba}.

To highlight the adaption of \textit{S.\,mutans} cells for the human oral cavity and salivary environments, we performed the exact same experiments with \textit{Staphylococcus carnosus} cells. \textit{S.\,carnosus} is an apathogenic representative of the genus \textit{Staphylococci} that is used in meat production and has no affinity for the oral cavity.

\section*{Materials and Methods}

\subsection*{Bacteria}

\textit{Streptococcus\,mutans} strain ATCC\,25175 was cultured from a deep-frozen stock solution on an agar medium selective for this species for three days\cite{Gold:1973gpba,Coykendall:1977by}. For every experiment, one colony from these plates (not older than  two weeks) was transferred into 5\,ml of sterile Todd Hewitt broth (THB) and cultured for 24\,hours at 37\,°C under agitation (150\,rpm). Afterwards, 40\,$\upmu$l of this solution were transferred into 4\,ml of fresh THB and cultered for another 16\,hours at 37\,°C and 150\,rpm. Finally, 1\,ml of this solution was washed three times with sterile PBS to remove any extracellular material and then stored at 4\,°C for not longer than 6\,hours.

For comparison, we used the apathogenic, non-oral species \textit{Staphylococcus\,carnosus} strain TM300\cite{Schleifer:1982vqbaca,Barriere:2001wabaca,Rosenstein:2009tjbaca,Rosenstein:2010tcbaca}.
These cells were grown from a deep-frozen stock solution on a blood agar plate for three days and a fresh plate was used not longer than two weeks. Before every experiment, one colony was suspended in 5\,ml tryptonic soy broth (TSB)  and cultured overnight at 37\,°C and 150\,rpm. From this solution, 40\,$\upmu$l were inoculated in 4\,ml fresh TSB and grown for another 2.5\,hours. Then, 1\,ml of this suspension was washed three times and afterwards stored at 4\,°C for not longer than 6\,hours.

\subsection*{Surfaces}

HAP pellets were produced by pressing and sintering of pure hydroxyapatite powder (Sigma Aldrich, Steinheim, Germany) resulting in an overall density of about 98\,\% of the density of a single crystal, following a standard procedure published earlier\cite{Loskill:2013jeba}. To increase their smoothness, pellets were treated with abrasive paper and polishing solutions of decreasing particle size (final polishing step with an diamond suspension of 30\,nm particle size). Subsequently, the samples were etched in an ultrasonic bath for 7\,s in sodium acetat buffer (pH\,4.5) to remove residues from the final polishing step. Finally, HAP pellets feature an rms roughness of less than 1\,nm, as determined by AFM. 

In preparation for every experiment, the HAP samples were cleaned for 5\,min in an ultrasonic bath in an aqueous solution of 1\%\,Mucasol (Merz Pharma, Frankfurt\,a.\,M., Germany). Afterwards, they were rinsed in an ultrasonic bath of pure deionized water (0.055\,$\frac{\upmu \text{S}}{\text{cm}}$, Thermo\,Fischer Scientific, Waltham, USA) for another 5\,min and dried in a stream of pure nitrogen.

\subsection*{Collecting of saliva}

On overall five different days, about 5\,ml each of saliva were collected from a volunteer with good oral health by chewing on parafilm (50\% paraffin and 50\% polyethylene),  and spitting into a sterile test tube. The volunteer refrained from eating and drinking (except for water) for one hour after brushing the theeth with normal  toothpaste. The saliva sample was collected 30\,min after renewed tooth brushing without tooth paste. Afterwards, the saliva samples were filtered first through a 2\,$\upmu$m and then through a 1\,$\upmu$m filter. Subsequently, they were frozen to -20\,°C. After the collection of all five samples, they were thawed, mixed together and again frozen to -20\,°C. For every experiment, a fresh sample was thawed and "vortexed" for 30\,s to ensure proper mixing of all saliva components.

\subsection*{Force/distance measurements}

\begin{figure}
	\centering
	\includegraphics[width=0.95\textwidth]{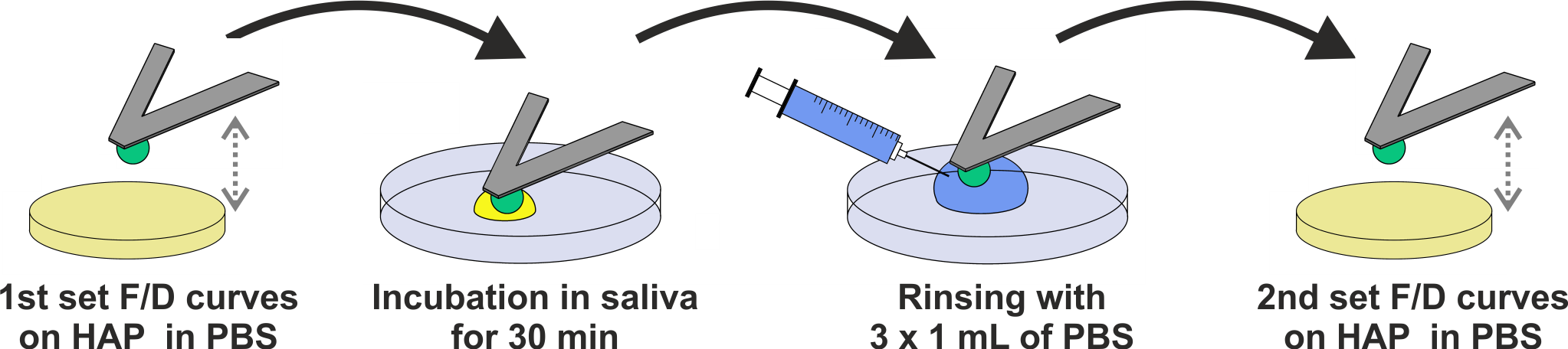}
	\caption{Scheme of the experimental procedure to inoculate a bacterial cell (green) with saliva.}
	\label{fig.howto}
\end{figure}

\begin{figure}[htbp]
	\centering
	\includegraphics[width=0.6\textwidth]{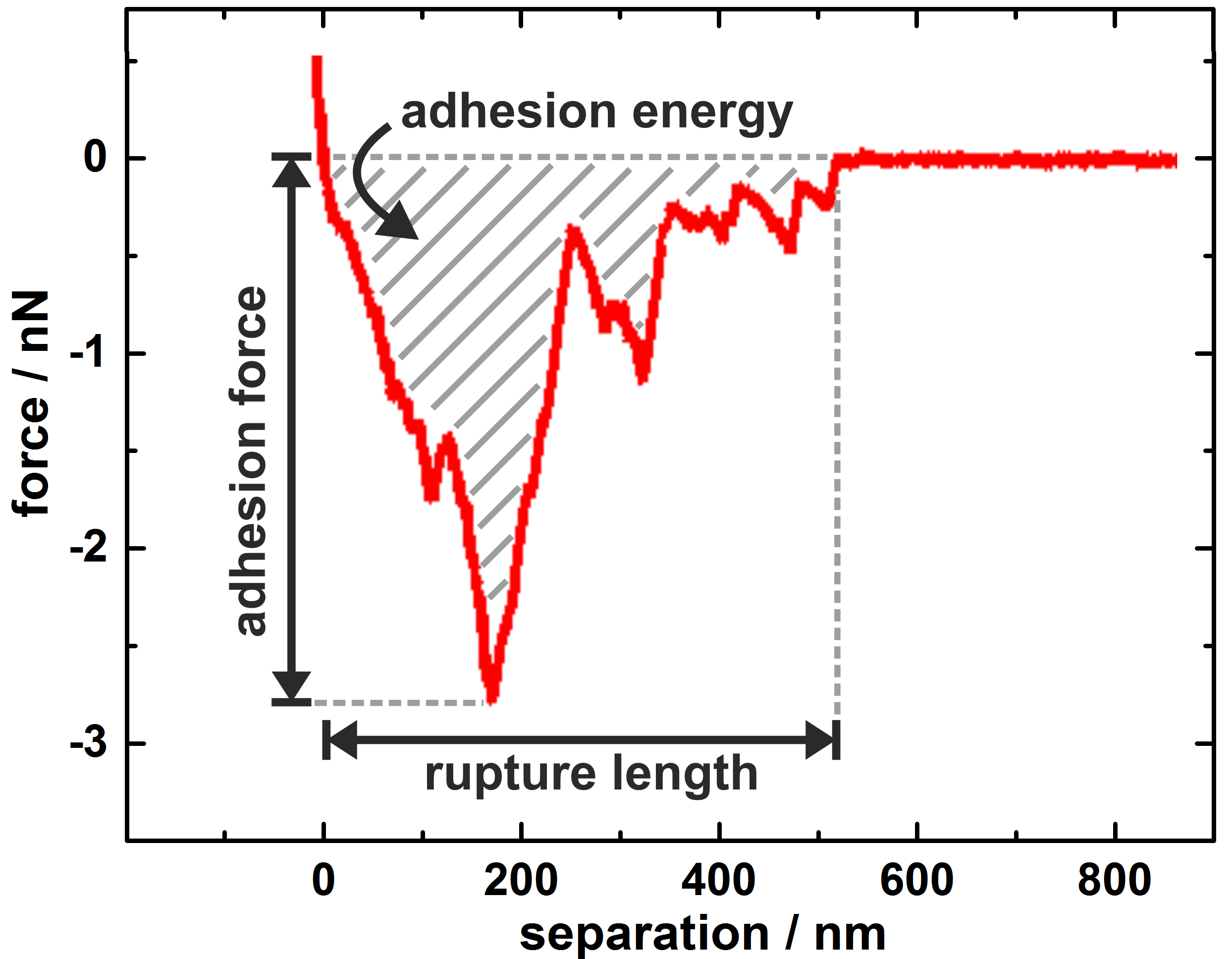}
	\caption{Retraction part of a typical force/distance curve, recorded with a single \textit{S.\,mutans} cell on HAP, displaying the main experimental measures.}
	\label{fig.fdexplanation}
\end{figure}

Force/distance measurements were performed on a Bioscope Catalyst (Bruker-Nano, Santa Barbara, USA) with  single bacterial cells immobilized on soft cantilevers, prepared by a method published earlier\,\cite{Thewes:2015heba}. We used tipless cantilevers (MLCT-0, Bruker) with a nominal spring constant of 0.03\,$\frac{\text{N}}{\text{m}}$ and a deflection sensitivity of 25\,$\frac{\text{nm}}{\text{V}}$. Cantilevers were calibrated before each set of experiments. 
The force trigger, which defines the maximum force with which the bacterial probe is pressed onto the substrate, was set to 300\,pN and the lateral distance between each single curve on the HAP surface was 1\,$\upmu$m. To test the influence of the binding kinetics, the force trigger can be hold constant for a certain time, called surface delay time (SD)\cite{Thewes:2014gqba,Zeng:2014bhba,Herman-Bausier:2015jjba}. We used SD of 0\,s, 2\,s and 5\,s. Thereby, 0\,s stands for a very short contact time of some ms\cite{Beaussart:2013g5ba}. For every bacterial cell, first three sets of 40  force/distance curves (one set for each surface delay time) were collected in phosphate buffered saline (PBS, pH\,7.3) on a bare HAP surface. Next, the bacterial cell - still immobilized on the cantilever - was covered with 50\,$\upmu$l of  filtered saliva for 30\,min. Then, the cantilever was washed three times  with  1\,ml of pure PBS each to remove possible leftovers of the saliva solution. Afterwards, the second three sets of force/distance curves were collected with the exact same parameters as before (see figure\,\ref{fig.howto}).
Subsequently, the measured force/distance curves were quantified in terms of adhesion force (minimal force during retraction), rupture length (distance between cell and HAP surface at which the last connection breaks) and  adhesion energy (area under the retraction curve), as shown in figure\,\ref{fig.fdexplanation}\cite{Thewes:2014gqba}.
Altogether, ten individual \textit{S.\,mutans} cells and five individual \textit{S.\,carnosus} cells were tested, and with one and the same bacterial cell, in sum 240 force/distance curves on the HAP surface were taken (120 before and 120 after inoculation). A possible deterioration due to the measurement can be excluded, since with increasing number of scans, no systematic change in the force curves (e.\,g. a decreasing adhesion) can be observed. This is in accordance with earlier studies \cite{Thewes:2014gqba}.

\section*{Results}

\begin{figure}[htbp]
	\centering
	\includegraphics[width=0.6\textwidth]{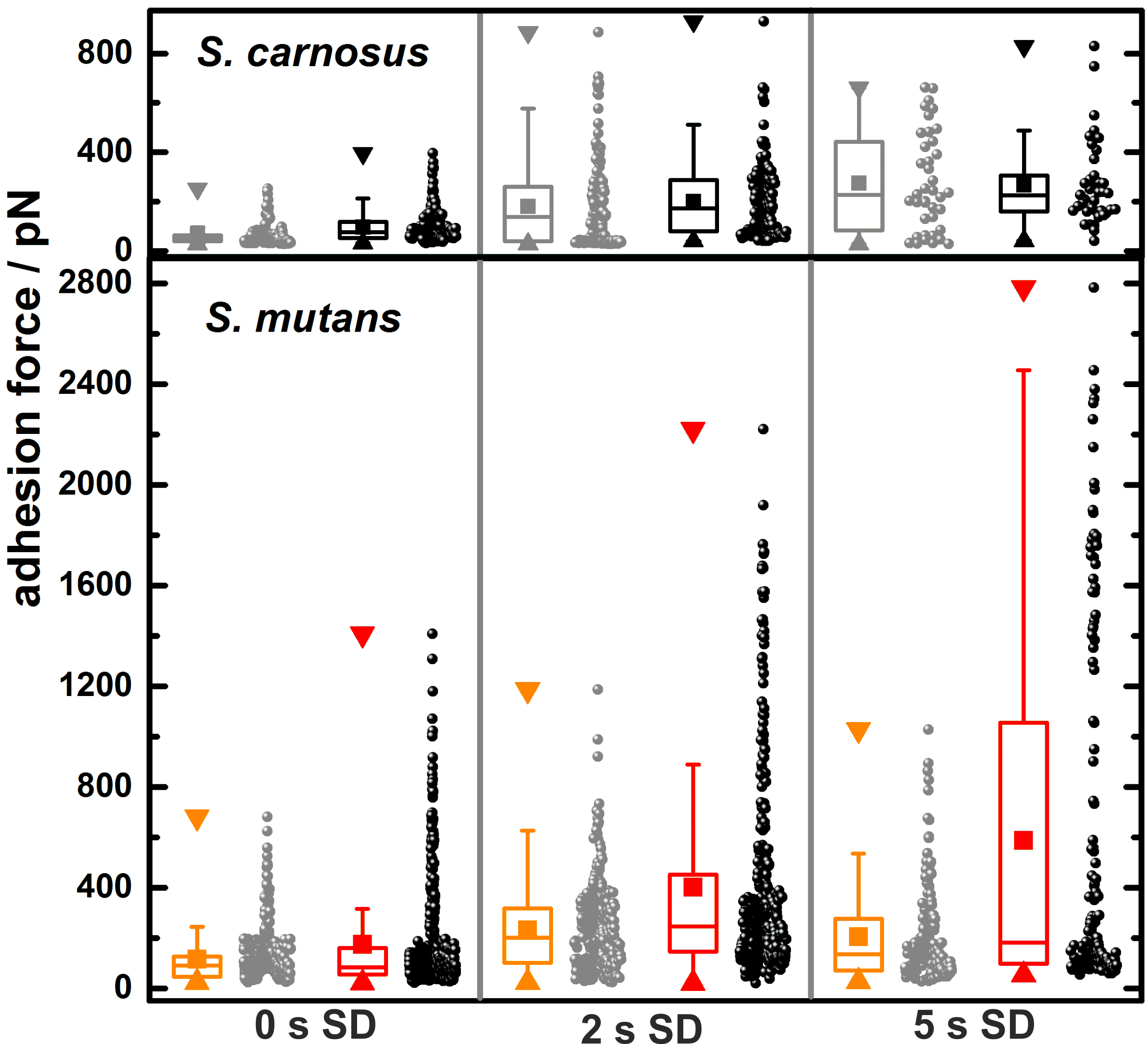}
	\caption{Adhesion forces of \textit{S.\,mutans} (lower panel) and \textit{S.\,carnosus} cells (upper panel) to HAP pellets before (light color) and after (dark color) saliva inoculation for different surface delay times (SD). For details of the box-and-whisker representation, see text.}
	\label{fig.force025_boxplot}
\end{figure}

Figure\,\ref{fig.force025_boxplot} depicts adhesion forces of ten individual \textit{S.\,mutans}  and five individual \textit{S.\,carnosus} cells before and after inoculation in saliva. Three different surface delay times have been tested to study the influence of contact time to adhesion strength. Overall, the adhesion forces cover a range of 0\,-\,2,800\,pN for \textit{S.\,mutans} and only 0\,-\,950\,pN for \textit{S.\,carnosus}.
Within the same surface delay panel of figure\,\ref{fig.force025_boxplot}, \textit{S.\,mutans} cells develops stronger adhesion forces after saliva inoculation, which is getting even more pronounced with longer surface delay times. Comparatively, \textit{S.\,carnosus} cells exhibit much smaller differences in adhesion force before and after inoculation. To display this trend, the results in figure\,\ref{fig.force025_boxplot} are shown as box-and-whisker plots, where the median is marked by the horizontal line in the box and the whiskers are defined as 1.5 times the extent of the interquartile range (IQR)\cite{Mcgill:1978tgba}.
These measures quantify what is already visible by eye: For both bacterial species, for $\text{SD} > 0\,\text{s}$, larger adhesion forces (median as well as average) are recorded, but the median is not significantly affected by saliva treatment. For \textit{S.\,mutans} cells, however, mean adhesion force, IQR as well as the whiskers gain significantly in size by saliva inoculation. In other words, in some force/distance measurements, the  treatment causes an especially enforced adhesion. 

\begin{figure}[htbp]
	\centering
	\includegraphics[width=0.6\textwidth]{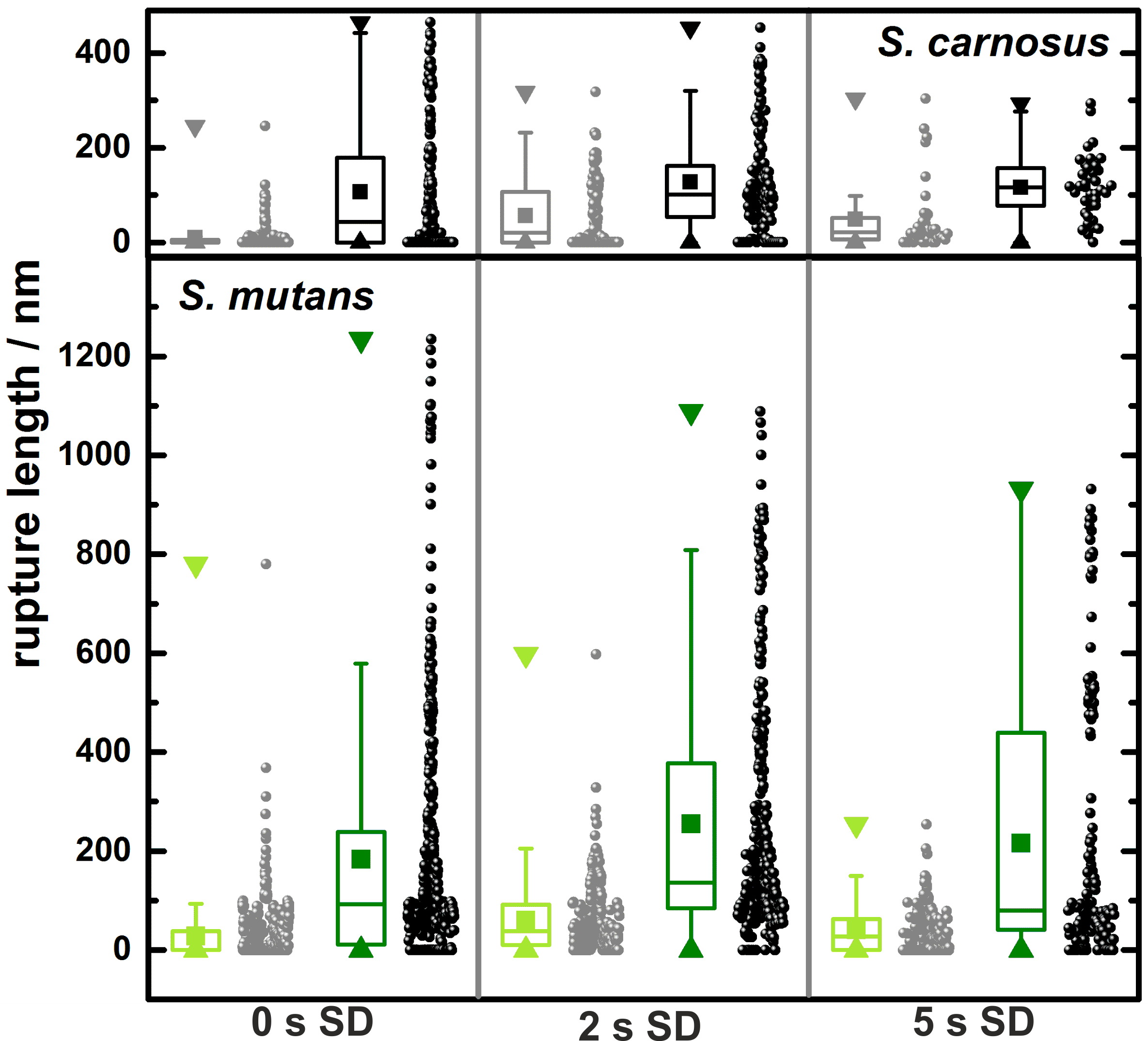}
	\caption{Rupture lengths for \textit{S.\,mutans} (lower panel) and \textit{S.\,carnosus} cells (upper panel) removed from HAP pellets before (light color) and after (dark color) saliva inoculation for different surface delay times (SD). For details of the box-and-whisker representation, see text.}
	\label{fig.distance025_boxplot}
\end{figure}

Similarly, as displayed in figure\,\ref{fig.distance025_boxplot}, the rupture lengths of \textit{S.\,mutans} cells are especially large after inoculation: No matter of the applied surface delay time, mean values of the rupture length increase by almost an order of magnitude after treatment. Before, rupture lengths are in the range of some tens of nanometers, whereas after the saliva treatment, they increase up to several hundreds of nanometers with a mean value of around 200\,nm and maxima up to 1,200\,nm. For \textit{S.\,carnosus} cells, rupture lengths on the HAP surface are in general smaller than for \textit{S.\,mutans} cells and the relative increase after saliva treatment is much smaller, only about a factor of two.

An even stronger effect of saliva treatment can be observed by examining the energy that is necessary to remove the entire bacterial cell from the surface. 
For \textit{S.\,mutans} cells, the mean value (as well as the IQR and the whiskers) is strongly increased: From a mean value of around 10,000\,$\text{k}_\text{B}\text{T}$ at 0\,s surface delay to a mean value of roughly 30,000\,$\text{k}_\text{B}\text{T}$ at 5\,s surface delay. Remarkably, at closer inspection, data points seem to develop a bimodal distribution at long surface delay times. Thereby, the median of the adhesion energy stays almost unchanged.

\begin{figure}[htbp]
	\centering
	\includegraphics[width=0.6\textwidth]{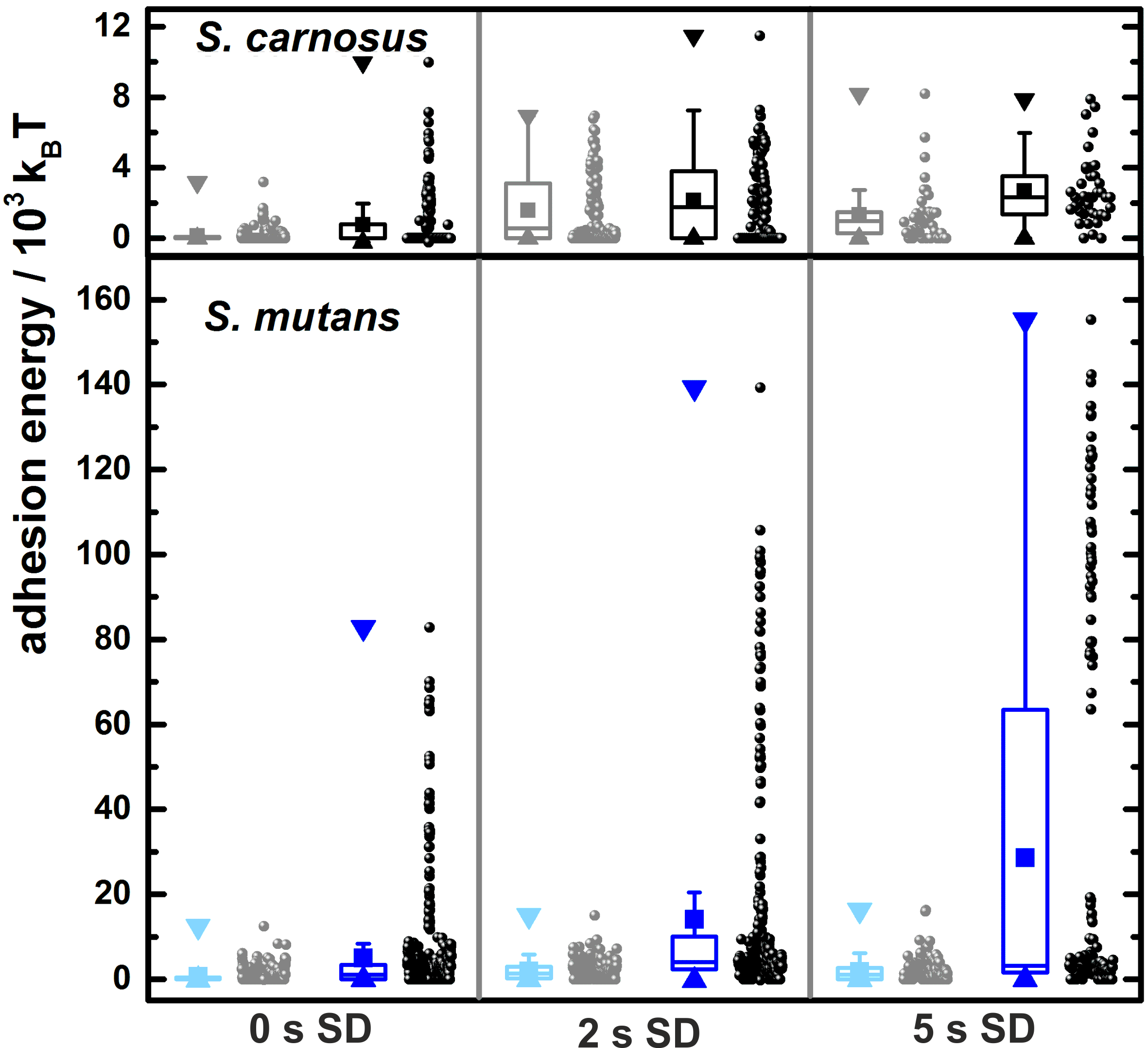}
	\caption{Adhesion energies of \textit{S.\,mutans} (lower panel) and \textit{S.\,carnosus} cells (upper panel) on HAP pellets before (light color) and after (dark color) saliva inoculation for different surface delay times (SD). For details of the box-and-whisker representation, see text. Note the fourfold stretched energy scale in the upper panel.}
	\label{fig.energy025_boxplot}
\end{figure}

\begin{figure}[htbp]
	\centering
	\includegraphics[width=0.6\textwidth]{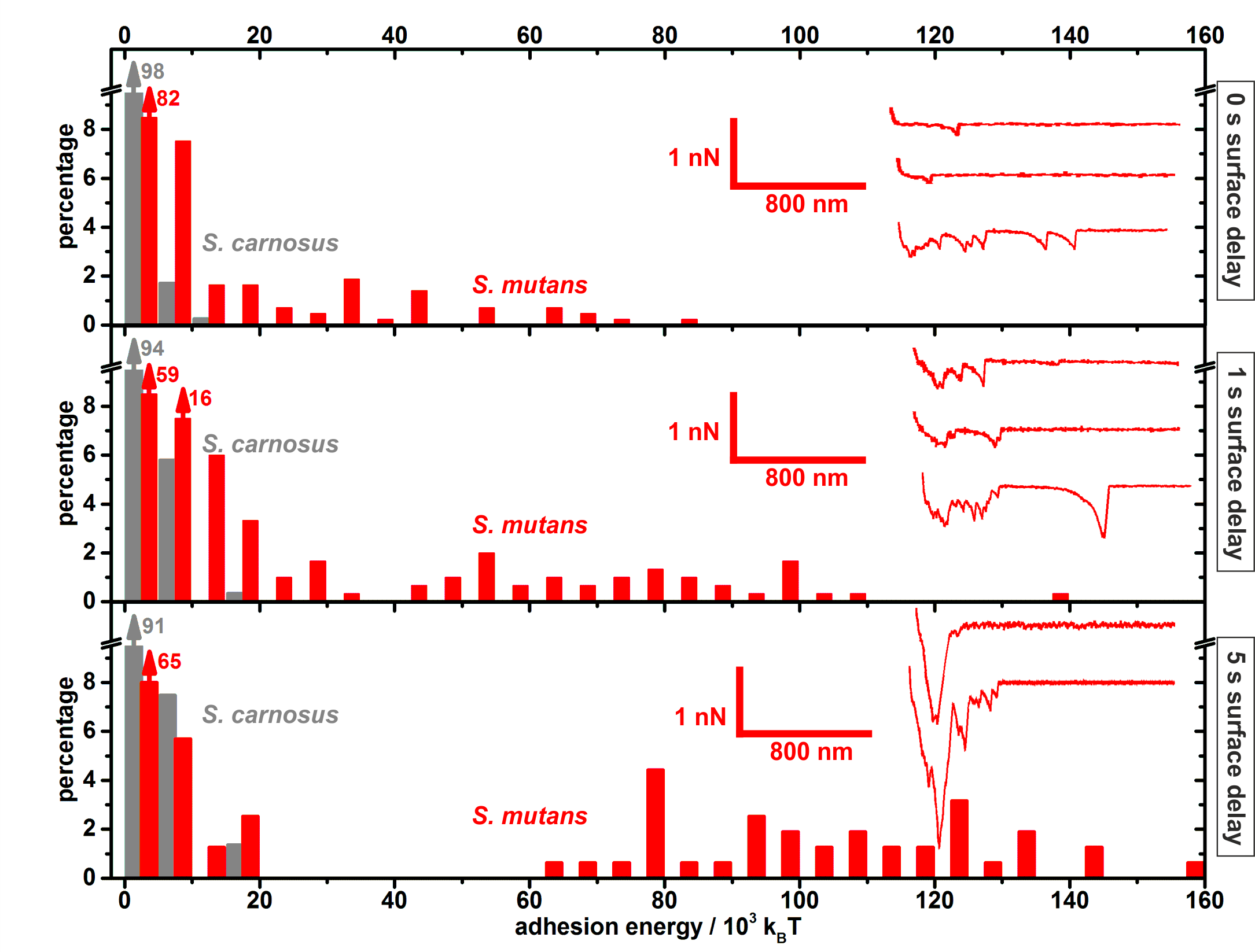}
	\caption{Adhesion energy histograms of \textit{S.\,mutans} (red) and \textit{S.\,carnosus} cells (gray) for different surface delay times after saliva inoculation. As insets, exemplary force/distance curves of \textit{S.\,mutans} cells are shown.}
	\label{fig.atcc_energy_histoS}
\end{figure}

To highlight this trend, figure\,\ref{fig.atcc_energy_histoS} shows histograms of  adhesion energies of \textit{S.\,mutans} 
and \textit{S.\,carnosus} 
cells on HAP after the inoculation in saliva for different surface delay times. With increasing surface delay, a second regime of large adhesion energies occurs for \textit{S.\,mutans} cells. Simultaneously, the portion of force/distance curves with adhesion energies below 20,000\,$\text{k}_\text{B}\text{T}$ decreases with increasing surface delay time. For SD\,$=$\,5\,s,  the mean value of the second regime in the histogram is located at adhesion energies of around 110,000\,$\text{k}_\text{B}\text{T}$. It is interesting to note that for \textit{S.\,mutans}, the different adhesion strengths become directly apparent in the way the shape of the force/distance curves changes with increasing surface delay times (eight curves are exemplarily shown as inset to figure\,\ref{fig.atcc_energy_histoS}): All curves display local minima and the deepest is taken as a measure for the adhesion force. However, for a surface delay of 5\,s, the deepest minimum in the retraction curve is much deeper than following local minima, whereas for curves with smaller surface delay times, all occurring local minima are in the same range of forces. 

For \textit{S.\,carnosus} cells, the scenario is completely different (see gray bars in figure\,\ref{fig.atcc_energy_histoS}): The adhesion energy is more than one order of magnitude smaller than for \textit{S.\,mutans} cells. 
Also, for all surface delay times, the energy histogram features only one regime and this is located at quite low adhesion energies of around 10,000\,$\text{k}_\text{B}\text{T}$.

\section*{Discussion}
Bacterial cells in the human mouth always run the risk of getting washed out, i.\,e. of getting swallowed. Therefore, the evolutionary success of mouth colonizing cells relies first of all on their ability to adhere in the oral environment. Here, we investigated the adhesion strength of cells of the mouth colonizing species \textit{S.\,mutans} to hydroxyapatite pellets before and after exposure of the cells to human saliva using AFM based single cell force spectroscopy. Our results demonstrate that the strength of adhesion between \textit{S.\,mutans} cells and hydroxyapatite increases significantly after exposure of the cells to saliva. In contrast, treating \textit{S.\,carnosus} cells (whose natural habitat is not the human mouth) in saliva does not increase the cells' adhesive strength to HAP pellets. Hence, \textit{S.\,mutans} cells exhibit a specific mechanism that enhances their adhesion in the human oral environment. This mechanism may be a result of the evolutionary adaption of this bacterial species to its natural habitat, the human oral cavity. Moreover, our study demonstrates that for a firm adhesion of \textit{S.\,mutans} cells to HAP surfaces, it is not necessary that SAG is present on the substratum, rather, even the exposure of the bacterial cell to a salivary environment is sufficient.

Open questions are why adhesion is enhanced by saliva inoculation and why \textit{S.\,mutans} cells are especially responsive to the treatment. The latter is not in the focus of our study, however, the former can be explained by the common notion how bacterial adhesion proceeds:
The adhesion process relies on the consecutive binding of bacterial cell wall macromolecules to a substratum\cite{Thewes:2015iuba,Song:2016gtba}. Since the binding strength of a single contact point cannot be altered by a saliva treatment of the bacterial cell, the only possibility to increase adhesion is to increase the number of contact points. So very likely, saliva treatment increases the number of macromolecules that tether to the surface. It remains open if i) the new cell wall macromolecules are certain salivary components that get linked to the bacterial cell wall via \textit{S.\,mutans} specific surface molecules (or domains), or if ii) \textit{S.\,mutans} produces additional cell wall macromolecules when exposed to its natural salivary environment.

Though the origin and the nature of these additional macromolecules remain unclear at this point, we can speculate about their properties, using results of this study: In force/distance curves, the adhesion force value is defined as the minimum force during retraction. The distance at which this point appears is related to the mechanical properties -- in particular the length –- of the contact forming macromolecules. In our measurements, the minimal force is usually located at separating distances of less than 200\,nm (see insets in figure\,\ref{fig.atcc_energy_histoS}). Hence, adhesion forces are dominated by rather short surface macromolecules that tether to a surface. These forces, however, are only slightly influenced by the saliva treatment. In contrast, the rupture lengths feature a strong increase after saliva inoculation. This is a clear hint that the additional macromolecules that contribute to the adhesion after saliva treatment are longer than the surface macromolecules responsible for adhesion before saliva treatment. The effect of saliva treatment has the strongest impact on the adhesion energy. This is the result of the combination of two effects, the slight increase in adhesion forces as well as the significantly larger rupture lengths.

The adhesion strength moreover increases with surface delay time. This can also be understood in light of the tethering bacterial cell wall macromolecules: For longer surface delay times, the effect of saliva treatment amplifies the increase in adhesion force because additional macromolecules have more time to find a suitable position to tether to the HAP surface. This means that for a longer surface delay time, more new macromolecules find such a binding site and therefore, the increase in adhesion force due to the salivary treatment even grows with longer surface delay times. In contrast, surface delay times do not cause longer rupture lengths, because in this case, it is sufficient that few (or in the extreme case only a single) additional, long macromolecules tether to the surface. For an adequately high number of additional, long macromolecules, already the initial contact between bacterial cell and surface leads to such a binding event and hence, an additional contact time does not have an influence.
 
It is also possible that for longer surface delay times not only more bonds originate but also existing bonds develop a stronger binding to the surface. This phenomenon, called bond strengthening, has been measured for \textit{Streptococci} as well as for \textit{Staphylococci}\cite{Mei:2008iuba, Boks:2008fiba}. However, in the present study, this effect is very likely not the primary reason for enhanced adhesion because bond strengthening usually appears when specific interactions between binding molecules of the cell and molecules on the surface are involved. In our case, though, the substratum is a bare, smooth hydroxyapatite surface, where no specific binding is expected. Furthermore, bond strengthening is usually observed for contact times notably longer than the 5\,s of this study. Yet, it was not possible with the present setup to apply longer surface delay times while keeping the force trigger constant. It shall be probed in the future if for much longer surface delay times, most data will fall into the second regime of the adhesion energy histogram.

\section*{Conclusions}

In this study, we have analyzed the adhesion properties of \textit{S.\,mutans} cells to hydroxyapatite surfaces. To monitor the adhesion process, atomic force microscope-based single cell spectroscopy was used on ultra smooth, high-density HAP pellets. It has been shown that adhesion force, rupture length and adhesion energy increase significantly when the cell has been inoculated in human saliva compared to adhesion without salivary treatment. Thereby, rupture length and adhesion energy are notably enlarged, which leads to the interpretation that especially long macromolecules contribute to this. These macromolecules either stem from the saliva or are produced by \textit{S.\,mutans} cells, stimulated by the contact to saliva.
By comparing identical measurements of \textit{S.\,mutans} and \textit{S.\,carnosus} cells, it has been shown that the adaption to a salivary environment is a particular property of \textit{S.\,mutans} cells and is far less pronounced for \textit{S.\,carnosus} cells.
Future studies may now further analyze saliva properties and identify salivary components that are responsible for this enhanced adhesion. That way, new pathways may open up in caries prevention by applying saliva-influencing agents after tooth-brushing.

\section*{Acknowledgement}
The study has been funded by the German Research Foundation (DFG) within the framework of SFB\,1027.

\clearpage
\newpage

\bibliography{saliva}

\end{document}